\pdfoutput=1
\documentclass{article}

\usepackage[utf8]{inputenc} 
\usepackage[T1]{fontenc}    
\usepackage{hyperref}       
\usepackage{url}            
\usepackage{booktabs}       
\usepackage{amsfonts}       
\usepackage{nicefrac}       
\usepackage{microtype}      

\usepackage{setspace} 
\usepackage[ruled,linesnumbered]{algorithm2e} 
\usepackage{float} 
\usepackage{adjustbox}
\usepackage{caption} 
\usepackage{cite} 
\usepackage{amsmath,amsfonts,amssymb} 
\usepackage[font=footnotesize,labelfont=bf]{caption}

\newcommand{\Tref}[1]{Table~\ref{#1}}

\newcommand{\fref}[1]{Fig.~\ref{#1}}
\newcommand{\Fref}[1]{Figure~\ref{#1}}

\newcommand{\Sref}[1]{Section~\ref{#1}}
\makeatletter
\DeclareRobustCommand\onedot{\futurelet\@let@token\@onedot}
\def\@onedot{\ifx\@let@token.\else.\null\fi\xspace}

\def\eg{\emph{e.g}\onedot} 
\def\ie{\emph{i.e}\onedot} 
 
\def\etc{\emph{etc}\onedot} 
 
\def\etal{\emph{et al}\onedot}
\makeatother
\usepackage{color}

\title{\large
	High-Performance Out-of-core Block Randomized
	Singular Value Decomposition on GPU}

\author{
	\small
	Yuechao Lu, Fumihiko Ino and Yasuyuki Matsushita \\
	\small
	Department of Computer Science,	Osaka University,	Osaka, Japan \\
	\small
	\texttt{\{yc-lu, ino, yasumat\}@ist.osaka-u.ac.jp} \\
}

\begin{document}
	
	\maketitle
	
	\begin{abstract}
		Fast computation of singular value decomposition (SVD) is of great interest in various machine learning tasks. Recently, SVD methods based on randomized linear algebra have shown significant speedup in this regime. This paper attempts to further accelerate the computation by harnessing a modern computing architecture, namely graphics processing unit (GPU), with the goal of processing large-scale data that may not fit in the GPU memory. It leads to a new block randomized algorithm that fully utilizes the power of GPUs and efficiently processes large-scale data in an out-of-core fashion. Our experiment shows that the proposed block randomized SVD (BRSVD)\footnote{Source code is available at \url{https://github.com/luyuechao}} method outperforms existing randomized SVD methods in terms of speed with retaining the same accuracy. We also show its application to convex robust principal component analysis, which shows significant speedup in computer vision applications.
	\end{abstract}
	
	\section{Introduction}
	\label{sec:introduction}
	
	The singular value decomposition (SVD) is an essential tool in machine learning, data analysis, and various scientific computing. Studies on improving the performance and numerical stability of SVD have been ongoing ever since its advent~\cite{golub1965calculating, larsen1998lanczos, frieze2004fast}, and have been successfully applied to applications such as principal component analysis (PCA)~\cite{pearson1901liii, hotelling1933analysis}. Recently, a randomized SVD (RSVD) method has been proposed to further accelerate SVD by exploiting the low-rank structure of data~\cite{halko2011finding, voronin2015rsvdpack} and now appears a method of choice for fast approximate SVD computation.
	
	While SVD has been made efficient in terms of its computational complexity based on these findings, a modern computing environment, where arithmetic operations are fast and highly parallelized but data communication is slow, requires a locality-aware method for fast SVD computation with efficient data access. Especially, for a large-scale data, the SVD computation cannot fully benefit from a fast level-3 basic linear algebra subprograms (BLAS3) computation, mainly general matrix-matrix product (GEMM), due to that the data may not fit in a single memory space and its computation pattern includes vast data accesses and communication between distinct memory hierarchy levels.
	
	Various modern applications in scientific computing and machine learning can be cast to many matrix operations~\cite{haidar2015framework}. That is to say, the application can be large-scale, but individual computing components are independent; therefore, they can be suitably mapped onto parallel execution units. As noted in~\cite{haidar2015framework,Haidar:2017:HCF:3038228.3038237}, it is important for such block operations to be executed on accelerators such as a graphics processing unit (GPU) without the host CPU's assistance, because performing non-trivial computations like BLAS routines on a host CPU may slow down the overall execution and reduce scalability. 
	
	The same applies to RSVD computation, and it is needed to develop a block algorithm of RSVD for its further acceleration. Motivated by this need, this paper considers redesigning the RSVD algorithm so that it performs efficiently with modern computation systems. In particular, we focus on the use of GPUs because of their relevance and high arithmetic computation capability. To attain high performance in RSVD on a GPU, there are two major bottlenecks described as follows that need to be accounted in practice.
	
	\noindent\textbf{CPU-GPU bandwidth bound.}
	GPU's arithmetic operations are becoming ever efficient, while communication costs are emerging as the bottleneck for a lot of applications not only in the multi-node distributed systems~\cite{christ2013communication} but also in single node CPU-GPU systems~\cite{jablin2011automatic}.
	
	
	
	\noindent\textbf{GPU memory capacity limit.} 
	Although a GPU has much higher arithmetic computation ability than a CPU (\eg, $5.3$ Tflop/s for an NVIDIA P100 GPU and $0.74$ Tflops/s for an Intel Xeon E5-2650v3 CPU in our system), its processing power for large-size data is greatly limited by the GPU memory capacity (\eg, 16GB for P100). Typically, \textit{divide-and-conquer} and \textit{pipelining}~\cite{sanders2010cuda,kirk2016programming} are employed as common approaches to overcome this limitation; however, these approaches may still be stalled by data dependency. 
	
	
	
	
	This paper presents a high-performance {\it out-of-core} (\ie, process data larger than GPU memory size) RSVD method for large matrices that do not fit in the GPU memory. Unlike previous approaches~\cite{yamazakirandom, mary2015performance} that mainly focus on arithmetic computation efficiency by exploiting multi-GPU or CPU/GPU hybrid systems, our method is designed to maximize the {\it pass-efficiency}~\cite{halko2011finding}, which measures the counts of data accesses, for handling large-scale data. The key idea is to fully utilize a block partitioning concept in the RSVD computation, and as a result we propose a {\it Block Randomized SVD} (BRSVD) method. The proposed BRSVD method reduces the number of data accesses by several times compared to a na\"{\i}ve implementation, effectively avoiding the data communication bottleneck. 
	
	To evaluate the efficiency of the proposed method, we compare the efficiency with an {\it in-core} (\ie, all working data can be held on a GPU memory) implementation, which is the performance upper-bound of RSVD. The result shows that, for small data, proposed method achieves the same level of efficiency compared with the in-core implementation without hurting the accuracy. For large data, the proposed out-of-core method achieves significant speedup in comparison to a conventional in-core method. 
	The attained efficiency of the approximate SVD computation is demonstrated by an application of the proposed method to robust principal component analysis (robust PCA)~\cite{candes2011robust}, which involves multiple SVD computations in its inner loop. The result shows the proposed method significantly outperforms the previous fast method~\cite{oh2015fast}.
	
	
	
	
	
	\section{Preliminaries} \label{sec:Preli}
	
	RSVD has been developed by Halko~\etal~\cite{halko2011finding} on top of the previous studies on randomized linear algebra~\cite{martinsson2006randomized,liberty2007randomized}. The randomization approach outperformed classical deterministic SVD methods in terms of speed with maintaining equivalent accuracy and robustness. 
	
	As described in~\cite{halko2011finding}, given a matrix $\mathbf{A} \in \mathbb{R}^{m \times n}$, an orthonormal basis $\mathbf{Q}$ can be constructed such that $\mathbf{A}\approx\mathbf{Q}\mathbf{Q}^{\top}\mathbf{A}$. The factorization (SVD, QR, \etc.) then can be efficiently computed using a relatively small {\it sketch matrix} $\mathbf{B}=\mathbf{Q}^{\top}\mathbf{A}$, when the basis matrix $\mathbf{Q}$ has few columns. In other words, when \hbox{$\mathrm{rank}(\mathbf{A}) = k \ll \min(m,n)$}, a small matrix $\mathbf{B}$ can be created and an SVD of the small matrix $\mathbf{B}$ reveals the SVD of the original matrix $\mathbf{A}$ as long as the range of the projector $\mathbf{Q}$ retains the action of the original matrix $\mathbf{A}$.
	
	The process of randomized factorization has two stages: (1) Constructing the basis $\mathbf{Q}$ with random projection of the original matrix $\mathbf{A}$, and (2) factorization of the small matrix \hbox{$\mathbf{B} = \mathbf{Q}^{\top} \mathbf{A}$} with a standard deterministic method. In stage (1), it is important to construct \hbox{$\mathbf{Q} = (\mathbf{q}^{(1)}, \ldots, \mathbf{q}^{(l)})$} such that it covers the range of $\mathbf{A}$. To achieve this, a random vector $\mathbf{\omega}$ can be used to form a sample vector $\mathbf{y}$ as 
	\begin{equation}
	\label{eqn:01}
	\mathbf{y}^{(i)} = \mathbf{A} \mathbf{\omega}^{(i)}, \quad i = 1, 2, \ldots, l,
	\end{equation}
	where $l = k + p$, and $p$ denotes the oversampling parameter. With $l$ samplings, a sample matrix $\mathbf{Y} = (\mathbf{y}^{(1)}, \ldots, \mathbf{y}^{(l)})$ can be constructed. In some cases, the singular spectrum of matrix $\mathbf{A}$ may decay slowly, power iteration is used to overcome this issue by projecting more information of $\mathbf{A}$ into the sample matrix $\mathbf{Y}$ so as to accelerate the spectrum decay. Afterward, the sample matrix $\mathbf{Y}$ is orthonormalized to create the basis of $\mathbf{Y}$ after power iterations by $\mathbf{A}\mathbf{A}^\top$:
	\begin{equation}
	\label{eqn:02}
	\mathbf{q}^{(i)} = \mathrm{orth} \left((\mathbf{A}\mathbf{A}^\top)^q\mathbf{y}^{(i)}\right),
	\end{equation}
	where the operator $\mathrm{orth}$ represents orthonormalization.
	In stage (2), matrix $\mathbf{B}$ is formed and factorized by conventional deterministic factorization methods. The RSVD algorithm can be summarized in Algorithm~\ref{alg:rsvd}. In stage (1), the random matrix \hbox{$\mathbf{\Omega} = \left(\mathbf{\omega}^{(1)}, \ldots, \mathbf{\omega}^{(l)}\right)$} is a standard Gaussian matrix of \textit{i.i.d} standard normal random variables with mean $0$ and variance $1$. In stage (2), a truncated SVD~\cite{golub2012matrix} is applied on the small matrix $\mathbf{B}$.
	
	\begin{algorithm}[t]
		\small
		\caption{Randomized SVD ~\cite{halko2011finding}}
		\label{alg:rsvd}
		\SetKwInOut{Input}{Input}
		\SetKwInOut{Output}{Output}
		\Input{ matrix $\mathbf{A} \in \mathbb{R}^{m \times n}$, target rank $k$, oversampling parameter $p$, and power iteration exponent $q$}
		\Output{ SVD of $\mathbf{A}$: matrices $\mathbf{U} \in \mathbb{R}^{m \times l}$, $\mathbf{\Sigma} \in \mathbb{R}^{l \times l}$, and $\mathbf{V}^{\top} \in \mathbb{R}^{l \times n}$}
		\BlankLine
		Generate a Gaussian matrix $\mathbf{\Omega} \in \mathbb{R}^{n \times l}$,
		where $l = k + p$.\label{alg:rsvd:1}
		
		$\mathbf{Y} = (\mathbf{A} \mathbf{A}^{\top})^q \mathbf{A} \mathbf{\Omega}$ \tcp*{sketch $\mathbf{A}$ and perform power iterations}
		
		$\mathbf{Q} = \mathrm{orthonormalize}(\mathbf{Y})$ \tcp*{form an orthonormal basis of $\mathbf{Y}$}
		
		$\mathbf{B} = \mathbf{Q}^{\top} \mathbf{A}$ \tcp*{form $\mathbf{B}$}
		
		$\left[ \widetilde{\mathbf{U}}, \mathbf{\Sigma}, \mathbf{V}^{\top} \right] = \mathrm{svd}(\mathbf{B})$ 
		\tcp*{truncated rank-$l$ SVD of $\mathbf{B}$} 
		
		$\mathbf{U} = \mathbf{Q}\widetilde{\mathbf{U}}$ \tcp*{form $\mathbf{U}$}
		
	\end{algorithm}

	There have been several efforts for accelerating the RSVD computation using GPUs. Yamazaki~\etal~\cite{yamazakirandom} proposed exploiting random sampling to update partial SVD on a hybrid CPU/GPU cluster. Their work showed that a random sampling algorithm achieves speedup of up to $14.1 \times$ compared to a standard deterministic SVD in a cluster environment. Voronin~\etal proposed a comprehensive randomized linear algebra library called RSVDPACK~\cite{voronin2015rsvdpack}. While effective, their GPU implementation is in-core, and the efficient computation can only be achieved when the data fits in the space of GPU memory. Ji~\etal~\cite{ji2014gpu} presented a GPU-accelerated implementation of RSVD for image compression. Their GPU/CPU hybrid implementation was about 6--7 times faster than the CPU version in their experiment. These studies pioneered utilizing GPU in accelerating RSVD and other randomized matrix decomposition algorithms, while they are all limited to either specialized purpose or memory capacity limit. Our work aim at providing a flexible and high-performance RSVD GPU solver for a wider range of applications that involve large data that cannot be accommodated in a GPU memory.
	
	
	Previous related works of RSVD on GPUs used GPU/CPU hybrid systems~\cite{yamazakirandom}. In these methods, only a part of RSVD computations, namely random number generation and GEMM, are performed on GPUs, but other computation steps, such as reduction of projected matrices, orthonormalization, and SVD are computed on CPUs, which results a large amount of CPU-GPU communications.
	The growth of GPU's arithmetic computation power is super-linear w.r.t. the growth of CPU-GPU communication bandwidth, and the CPU-GPU bandwidth bound emerges as a new bottleneck for various applications~\cite{Haidar:2017:HCF:3038228.3038237}. Our method avoids this bottleneck by maximizing the pass-efficiency with a block partitioning approach and performing all the RSVD computation on a GPU.
	
	\section{Proposed Method: Block Randomized SVD (BRSVD) on a GPU}

	\begin{figure}[t]
		\centering
		\includegraphics[width=1\hsize]{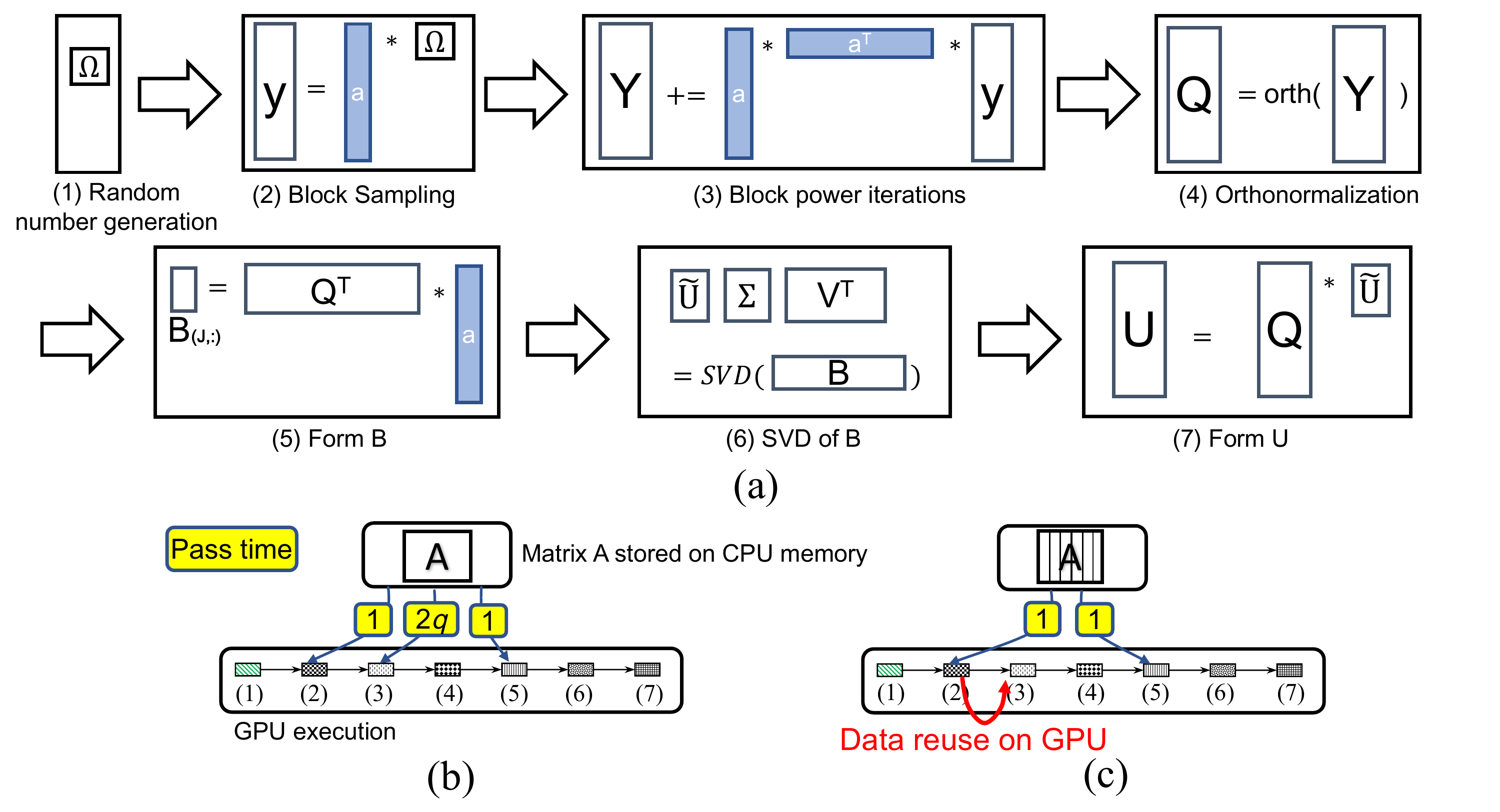}
		\caption{(a) Proposed block randomized SVD (BRSVD) method. A column block $\mathbf{a}$ is reused on a GPU in the RSVD computation pipeline. \textit{Pass-efficiency} comparison of (b) na\"{\i}ve implementation and (c) proposed BRSVD illustrates that our method reduces data accesses for efficient computation.}
		\label{fig:block}
	\end{figure}
	
	
	The proposed BRSVD method partitions the input matrix $\mathbf{A} \in \mathbb{R}^{m \times n}$ into column blocks that consists of a set of columns and sequentially transfers them to GPU. Subsequently, a sample matrix $\mathbf{Y} \in \mathbb{R}^{m \times l}$ is constructed in a gradual manner by capitalizing on the fact that matrix product can be naturally divided as:
	\begin{eqnarray}
	\mathbf{Y} = \mathbf{A} \mathbf{\Omega} = \sum_{i,j} \mathbf{A}_{(:,j)} \mathbf{\Omega}_{(i,:)},
	\end{eqnarray}
	where the subscript indicates matrix elements; $(i,:)$ and $(:,j)$ designate the $i$-th row and $j$-th column, respectively.
	For each column block $\mathbf{A}_{(:,J)} \in \mathbb{R}^{m \times n'}$ where $J$ is a list of column indices ($|J| = n'$), a Gaussian matrix $\mathbf{\Omega}_{j} \in \mathbb{R}^{n' \times l}$ is drawn on the GPU to sketch the column block $\mathbf{A}_{(:,J)}$ by $\mathbf{A}_{(:,J)} \mathbf{\Omega}_{j}$. The resulting matrix is further refined via a power method (with exponent $q$) reusing the transferred column block $\mathbf{A}_{(:,J)}$, and the sample matrix $\mathbf{Y} \in \mathbb{R}^{m \times l}$ is updated as
	\begin{eqnarray}
	\mathbf{Y} \leftarrow \mathbf{Y} + \left(\mathbf{A}_{(:,J)} \mathbf{A}_{(:,J)}^\top \right)^q \mathbf{A}_{(:,J)} \mathbf{\Omega}_{j}.
	\end{eqnarray}
	After each update of $\mathbf{Y}$, the transfered column block $\mathbf{A}_{(:,J)}$ is discarded from the GPU for avoiding memory overflow.
	
	
	\begin{algorithm}[t]
		\small
		\caption{Proposed method: RSVD by block column sampling}
		\label{alg:colSampling}
		\SetKwInOut{Input}{Input}
		\SetKwInOut{Output}{Output}
		\Input{ matrix $\mathbf{A}  \in \mathbb{R}^{m \times n}$, target rank $k$, oversampling parameter $p$, power iteration exponent $q$, and partition number $s$}
		
		\Output{ SVD of $\mathbf{A}$: matrix $\mathbf{U} \in \mathbb{R}^{m \times l}$, $\mathbf{\Sigma} \in \mathbb{R}^{l \times l}$, and $\mathbf{V}^\top \in \mathbb{R}^{l \times n}$}
		\BlankLine
		$n' = \lceil n/s \rceil;\ \ l = k+p;\ \ \mathbf{Y} = \mathbf{0}_{m \times l}$;
		
		\For(\tcp*[f]{$J$ denotes index set $jn':(j+1)n'-1$} ) {$j \gets 0$ {\rm \bf to} $s-1$}{
			Generate a Gaussian matrix $\mathbf{\Omega}_j \in \mathbb{R}^{n' \times l}$;
			
			$\mathbf{Y} \leftarrow \mathbf{Y} + \left(\mathbf{A}_{(:,J)} {\mathbf{A}_{(:,J)}}^{\top} \right)^q {\mathbf{A}_{(:,J)}} \mathbf{\Omega}_j$ \tcp*{sketch $\mathbf{A}_{(:,J)}$ and do power iterations} 
		}
		$\mathbf{Q}_{m \times l} = \mathrm{orthonormalize}(\mathbf{Y})$ \tcp*{orthonormalization by CAQR}
		free $\mathbf{Y}$; \ $\mathbf{B} = \mathbf{0}_{l \times n}$;
		
		\For{$j \gets 0$ {\rm \bf to} $s-1$}{
			
			$\mathbf{B}_{(:,J)} \leftarrow \mathbf{Q}^\top \mathbf{A}_{(:,J)}$ \tcp*{form $\mathbf{B}$}
		}
		$\left[ \widetilde{\mathbf{U}}, \mathbf{\Sigma}, \mathbf{V}^{\top} \right] = \mathrm{svd}(\mathbf{B})$ 
		\tcp*{truncated rank-$l$ SVD of $\mathbf{B}$} 
		
		$\mathbf{U} = \mathbf{Q}\widetilde{\mathbf{U}}$ \tcp*{form $\mathbf{U}$}
	\end{algorithm}
	
	Once the sample matrix $\mathbf{Y}$ is created, its orthonormalized basis $\mathbf{Q} \in \mathbb{R}^{m \times l}$ is constructed on the GPU. Using the basis $\mathbf{Q}$, a small core matrix $\mathbf{B} \in \mathbb{R}^{l \times n}$ is computed using the column block $\mathbf{A}_{(:,J)}$ that is once again transferred from the CPU memory as
	\begin{eqnarray}
	\mathbf{B}_{(J,:)} \leftarrow \mathbf{Q}^\top \mathbf{A}_{(:,J)}.
	\end{eqnarray}
	Finally, an SVD of the small matrix $\mathbf{B}$ is performed on the GPU to yield its decomposition $\widetilde{\mathbf{U}}, \mathbf{\Sigma}, \mathbf{V}^\top$, and by reprojecting the obtained basis $\widetilde{\mathbf{U}}$ by $\mathbf{Q}$, the left singular vectors $\mathbf{U} = \mathbf{Q} \widetilde{\mathbf{U}}$ of the input matrix $\mathbf{A}$ can be obtained.
	\Fref{fig:block} (a) summarizes the overall pipeline of the proposed method.
	
	While a na\"{\i}ve implementation requires $2(q+1)$ times of data accesses to the input matrix $\mathbf{A}$, the proposed BRSVD method only requires twice of data transfers because it reuses the column blocks $\mathbf{A}_{(:,J)}$ on the GPU memory (see \fref{fig:block} (b) and (c)). As we will see later in the experiment, this reduction of data transfer significantly improves the efficiency of RSVD computation for large matrices. For now, let us look at the efficiency analysis summarized \Tref{tb:flops}. It compares the computation and communication costs of the proposed method with a na\"{\i}ve method. In the table, \#flops refers to the arithmetic computation cost in floating point operations. \#words refers to the communication cost between CPU and GPU memory. While the \#flops remains the same in both approaches, our BRSVD method significantly reduces the communication cost \#words from $\mathcal{O}(mnl+(m+n)l^2)$ to $\mathcal{O}(m(n+l))$.
	
	\paragraph{Implementation details.}
	Here we describe implementation details that will be needed to reproduce the work. \\
	\noindent\textit{Sampling and power iteration}: For generating Gaussian random matrices $\mathbf{\Omega}_J$ on GPU, we have used cuRAND library~\cite{curand}. The random number generation is performed in parallel with transferring column blocks $\mathbf{A}_{(:,J)}$ and sampling of the column blocks. 
	The GEMM calculation sequence in Line 4 of Algorithm~\ref{alg:colSampling} is reversed from right to left based on the associative law of matrix multiplication so as to avoid generating a large projection matrix of size $m \times m$ in the process:
	\begin{equation}
	\label{eqn:04}
	\mathbf{Y} \leftarrow \mathbf{Y} +  \overleftarrow{\underbrace{\mathbf{A}_{(:,J)} {\mathbf{A}_{(:,J)}}^{\top}\cdots \mathbf{A}_{(:,J)} {\mathbf{A}_{(:,J)}}^{\top}}_{q \quad (\mathrm{power~iteration})} {\mathbf{A}_{(:,J)}} \mathbf{\Omega}_J},
	\end{equation}
	in which the long arrow on the top represents the order of matrix multiplication.
	
	\noindent\textit{Orthonormalization}: To orthonormalize the sample matrix $\mathbf{Y}$, instead of using a classical Gram-Schmidt (CGS)~\cite{yamazakirandom} or Cholesky QR (CholQR)~\cite{mary2015performance}, we use a communication-avoiding QR (CAQR) factorization method proposed by Demmel~\etal~\cite{demmel2012communication,anderson2011communication}. CAQR has a lower communication cost compared to Householder QR~\cite{golub2012matrix} and achieves fewer data accesses between GPU kernels and memory. In our test, it runs roughly $1.5\times$ faster than a {\tt orgqr()} routine provided in MAGMA 2.2 library~\cite{tnld10}. In addition, since CAQR is built on the block Householder QR~\cite{demmel2012communication}, it has intrinsically higher numerical stability than CGS and CholQR. 
	
	\noindent\textit{SVD}: For an SVD on the small matrix $\mathbf{B}$ on the GPU, we compared {\tt gesvd()} routines provided by MAGMA and cuSOLVER~\cite{cusolver} library and found that there were not much performance difference in terms of both speed and accuracy. We therefore chose cuSOLVER included in the CUDA library to keep the implementation simple and portable.
	
	

	\begin{spacing}{1}
		
		\begin{table}[t]
			\small
			\centering
			\captionof{table}{Computation and communication costs comparison. \#flops refers to the arithmetic computation cost in floating point operations, and \#words indicates the communication cost between CPU and GPU memory. Line \# indicates the corresponding operation blocks in Algorithm~\ref{alg:colSampling}.} \label{tb:flops}
			\begin{adjustbox}{width=1\textwidth}
				\begin{tabular}{l|l||l|l|l} 
					& line \#& \#flops & \#words (na\"{\i}ve)  & \#words (proposed)\\
					\hline
					(1) Random number generation& 3 & $\mathcal{O}(nl)$ & $\mathcal{O}(nl)$  
					& $0$  \\
					
					(2) Sampling & 4 & $\mathcal{O}(mnl)$ & $\mathcal{O}(mnl)$ 
					& $\mathcal{O}(mn) $\\
					
					(3) Power Iterations & 4 & $\mathcal{O}(mnlq)$ & $\mathcal{O}(mnlq)$ 
					& $0$ \\
					
					(4) Orthonormalization & 6 & $\mathcal{O}(ml^2)$ & $\mathcal{O}(ml)$ 
					& $0$ \\ 
					
					(5) Form $B$ & 9 & $\mathcal{O}(mnl)$ & $\mathcal{O}(mnl)$ 
					& $\mathcal{O}(mn)$ \\
					
					(6) SVD & 11& $\mathcal{O}(nl^2)$  &$\mathcal{O}(nl)$ 
					&$0$ \\
					(7) Form $U$ &12 & $\mathcal{O}(ml^2)$ & $\mathcal{O}(ml^2)$ 
					& $\mathcal{O}(ml)$  \\ 
					\hline
					\ \ \ \ \ \ \ \ \ Total& & $\mathcal{O}(mnl+(m+n)l^2)$ & $\mathcal{O}(mnl+(m+n)l^2)$
					& $\mathcal{O}(m(n+l))$ \\
				\end{tabular}
			\end{adjustbox}
		\end{table}
	\end{spacing}

\section{Experiments} \label{sec:experiments}
We now show empirical performance comparison of the proposed BRSVD method with other three RSVD implementations listed below. All the implementations are carefully optimized so as to yield best performance in each setting.
\begin{enumerate} 
	\item \textbf{na\"{\i}ve by cuBLAS-XT}: cuBLAS-XT~\cite{cublas} is a BLAS3 routines that can process data larger than the GPU memory. It frees users from dealing with GPU memory allocation and CPU-GPU communication; however, it cannot control the reuse of the transferred data. This implements Algorithm~\ref{alg:rsvd} using the cuBLAS-XT package.
	\item \textbf{na\"{\i}ve by CPU}: This is a straightforward implementation of Algorithm~\ref{alg:rsvd} on CPUs, with which all the working data is processed on CPUs.
	\item \textbf{in-core on GPU}: This is an in-core GPU implementation, with which all the working data is held on the GPU memory. Since there is no data communication between CPU-GPU, we can expect the fastest processing speed. However, it only works for a small scale data that can well fit in the GPU memory. 
\end{enumerate}

\paragraph{Experiment environment and setup.}

For evaluating the GPU implementations, we used NVIDIA P100 (Pascal) GPU with 16GB memory. A P100 was connected to the host via PCIe 3.0 interface. The theoretical peak in double precision is $5.3$ Tflop/s ($=2\times1.48$GHz$\times 3584$-core). Although P100 can access its own memory at $732$ GB/s, the speed of CPU to GPU data transfer is limited to $15.8$ GB/s at maximum. cuBLAS 8.0 and cuSLOVER 8.0 were used for BLAS and solver routines, respectively. 
For assessing the CPU implementation, we use a system equipped with two-socket Intel $10$-core Xeon E5-2650v3 (Haswell) processors with 128GB DDR4-2133 memory. The theoretical peak in double precision is $0.736$ Tflop/s ($=16\times2.3$GHz$\times10$-core$\times$2-socket) for two CPUs. Intel MKL (Math Kernel Library)~\cite{mkl} 11.3.1 is used for BLAS and solver routines.

Regarding the data, we generated matrices with various shapes and sizes with different ranks. A low-rank input matrix $\mathbf{A} \in \mathbb{R}^{m \times n}$ with rank-$k$ was created by a product of two low-dimensional matrices $\mathbf{A}_l \in \mathbb{R}^{m \times k}$ and $\mathbf{A}_r \in \mathbb{R}^{k \times n}$ that were both random Gaussian matrices. About the block size selection, we maximizing the block size by querying available memory size at runtime.


	
\paragraph{Results.}\label{subsec:perfm}
	
\begin{figure}[tb]
	\centering
	\includegraphics[width=1\hsize]{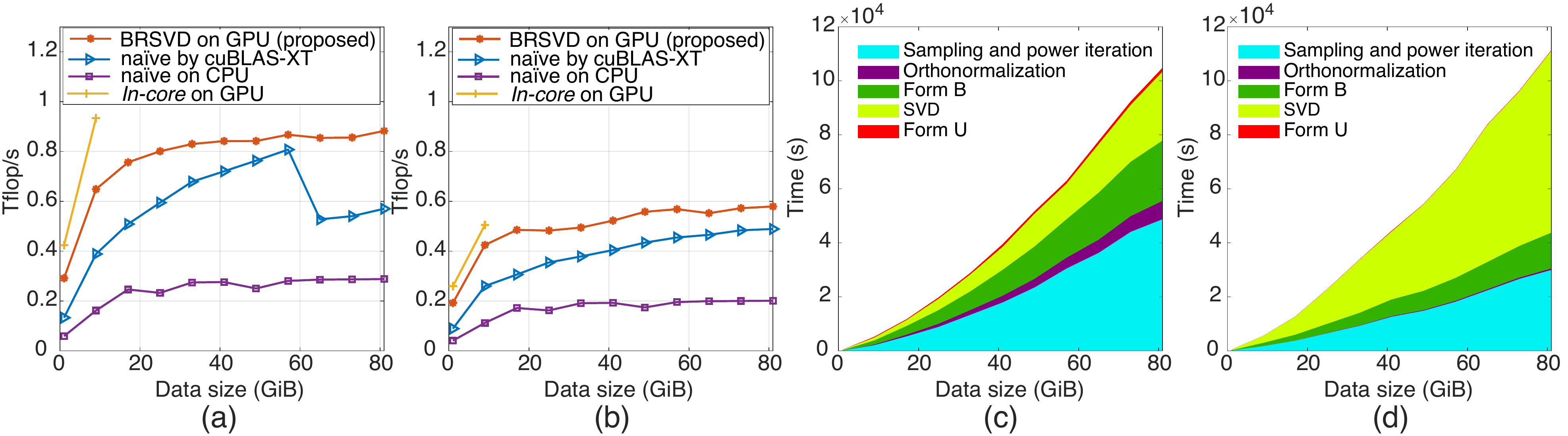}
	\caption{Performance comparison for different data sizes in double precision. (a) and (b) show overall performance of tall-skinny matrices ($m:n:k=1024:32:1$) and square ($m:n:k=256:256:1$) ones in Tflops/s. (c) and (d) show the time breakdown of proposed method for tall-skinny and square matrices.}
	\label{fig:tflops}
\end{figure}
	
\Fref{fig:tflops} shows experimental results on tall-skinny (a, c) and square (b, d) matrices individually. In each experiment, the ratio of matrix dimensions $(m, n)$ and rank $k$ are fixed, and the performance is measured by varying the size of input data. For the attained performance, all measurements include CPU-GPU data transfer time.
	
The experimental results show good scalability for our proposed block sampling method compared to the na\"{\i}ve by cuBLAS-XT implementation. Interestingly, even compared with the in-core on GPU implementation in \fref{fig:tflops} (a) and (b), there is a only fraction of performance penalty. There is a performance drop for the na\"{\i}ve by cuBLAS-XT implementation in processing data larger than $60$ GB. We speculate that it may be caused by that the input data is separated on the physical memory of two CPUs and the data routing has not been optimized in cuBLAS-XT yet.
	
\Fref{fig:tflops} (c) shows that the GEMM dominates the running time for tall-skinny matrices. For square matrices, due to the increased portion of SVD computation which renders low flop/s for both CPU and GPU, the overall performance in \fref{fig:tflops} (d) is almost reduced by half compared with tall-skinny ones.
	
Concerning decomposition accuracy we use \hbox{$\| \mathbf{A} - \mathbf{U} \mathbf{\Sigma} \mathbf{V}^{\top}\|_F / \| \mathbf{A}\|_F$} to quantify the approximation error. For double and single precision routines, we achieve the same level of error ($10^{-15}$ for double and $10^{-7}$ for single precision) as the CPU implementation based on MKL. We leave out the results for limited space.



\section{Application to Robust PCA via Inexact Augmented Lagrange Multiplier} \label{sec:app}
 
Robust PCA (RPCA) is a method used to recover a low-rank matrix with an unknown fraction of data corruption~\cite{candes2011robust}. It is widely used in applications such as computer vision, data mining, and bioinformatics. While applications like neural networks benefit a lot by accelerating on GPUs, RPCA did not benefit much from this emerging parallel computing approach. Mainly this is caused by the low flop/s of SVD on GPUs, which hinders adopting GPUs to accelerate RPCA.
 
 RPCA tries to separate the sparse corruptions $\mathbf{S}$ from the original data $\mathbf{M}$, so that a low-rank matrix $\mathbf{L}$ can be obtained as $\mathbf{L} = \mathbf{M} - \mathbf{S}$. The formulation of this problem can be written as:
 \begin{equation}
 \label{eqn:03}
 \min_{\mathbf{L}, \mathbf{S}}{||\mathbf{L}||_{*}+\lambda||\mathbf{S}||_{1}}  \ \ \ \mathrm{subject} \ \mathrm{to}  \ \ \ \mathbf{M} = \mathbf{L} + \mathbf{S}, 
 \end{equation}
 where $||\cdot||_{*}$ and $||\cdot||_1$ denote the nuclear norm and $\ell_1$ norm of a matrix, respectively, and $\lambda$ is a positive weighting parameter that is usually set to $1/\sqrt{\max(m,n)}$. Various solvers have been proposed for this convex optimization problem, and recent solution methods drastically improved the efficiency~\cite{yuan2009sparse, lin2009fast, lin2010augmented}.
 

 
In our test, we find the Inexact Augmented Lagrange Multiplier (IALM) method~\cite{lin2010augmented} is fast and stable compared to others that are even newer. In a similar manner to~\cite{oh2015fast}, we have used our BRSVD for the SVD computation block that appears in the inner loop of the RPCA with IALM, resulting in Algorithm~\ref{alg:rpca}. 
The shrinkage operator ${\mathcal{S}}_{{\varepsilon}}$ in line 4 of Algorithm~\ref{alg:rpca} is defined as
 \begin{equation}
 {\mathcal{S}}_{\varepsilon}=\left\{\begin{matrix}
 x-{\varepsilon},\ \ \textup{if}\ x>{\varepsilon},  \\ 
 \ \ \ x+{\varepsilon},\ \ \textup{if}\ x<{-\varepsilon}, \\ 
 \ 0, \ \ \ \ \ \ \ \ \textup{otherwise.}
 \end{matrix}\right.
 \end{equation} 
We applied Algorithm~\ref{alg:rpca} to two computer vision applications: shadow removal and background subtraction. The purpose is mainly assessing the computation speed in these practical applications, since the accuracy is almost equivalent.

 \begin{spacing}{1}
 	\begin{algorithm}[H]
 		\caption{RPCA via the Inexact ALM Method by RSVD}
 		\label{alg:rpca}
 		\SetKwInOut{Input}{Input}
 		\SetKwInOut{Output}{Output}
 		\Input{ matrix $\mathbf{M} \in \mathbb{R}^{m \times n}$, target rank $k$, oversampling parameter $p$, power iteration exponent $q$, convergence condition $\mathrm{tol}$.}
 		\Output{ low-rank matrix $\mathbf{L}\in \mathbb{R}^{m \times n}$, and sparse matrix $\mathbf{S}\in \mathbb{R}^{m \times n}$}
 		\BlankLine
 		$\mathbf{Y}_0 = \mathbf{M} \ / \ \max(||\mathbf{M}||_2, \ \lambda^{-1}||\mathbf{M}||_{\infty});\  \mathbf{E}_0 = \mathbf{0}; \mu_{0} > 0;\  \rho > 1;\  i = 0$
 		
 		\While{ TRUE~ }{
 			$[\mathbf{U}, \mathbf{\Sigma}, \mathbf{V}^T] = \mathrm{rsvd}(\mathbf{M} - \mathbf{S}_{i} + {\mu_i}^{-1} \mathbf{Y}_i \ ,\ k, \ p, \ q)$;
 			
 			$\mathbf{L}_{i+1} = \mathbf{U}{\mathcal{S}}_{{\mu_i}^{-1}}[\mathbf{\Sigma}] \mathbf{V}^T$\tcp*{shrinkage operation on singular values $\mathbf{\Sigma}$}
 			
 			$\mathbf{S}_{i+1} = \mathcal{S}_{{\lambda\mu_i}^{-1}}[\mathbf{M} - \mathbf{L}_{i+1} + {\mu}^{-1}\mathbf{Y}_{i}]$;
 			
 			$\mathbf{Y}_{i+1} = \mathbf{Y}_{i} + {\mu}_i(\mathbf{M} - \mathbf{L}_{i+1} - \mathbf{S}_{i+1})$;
 			
 			$\mathbf{Z}_{i+1} = \mathbf{M} - \mathbf{L}_{i+1} - \mathbf{S}_{i+1}$;
 			
 			\textbf{if}$\ (\ ||\mathbf{Z}||_F \ / \ ||\mathbf{M}||_F < \mathrm{tol}\ ) \ $ \textbf{break}\tcp*{evaluate convergence}
 			
 			$\mu_{i+1}=\rho\mu_i\ ; \ \ i= i+1$;  
 		}
 		
 	\end{algorithm}
 \end{spacing}
 
 \begin{figure}[t]
 	\centering
 	\includegraphics[width=1\hsize]{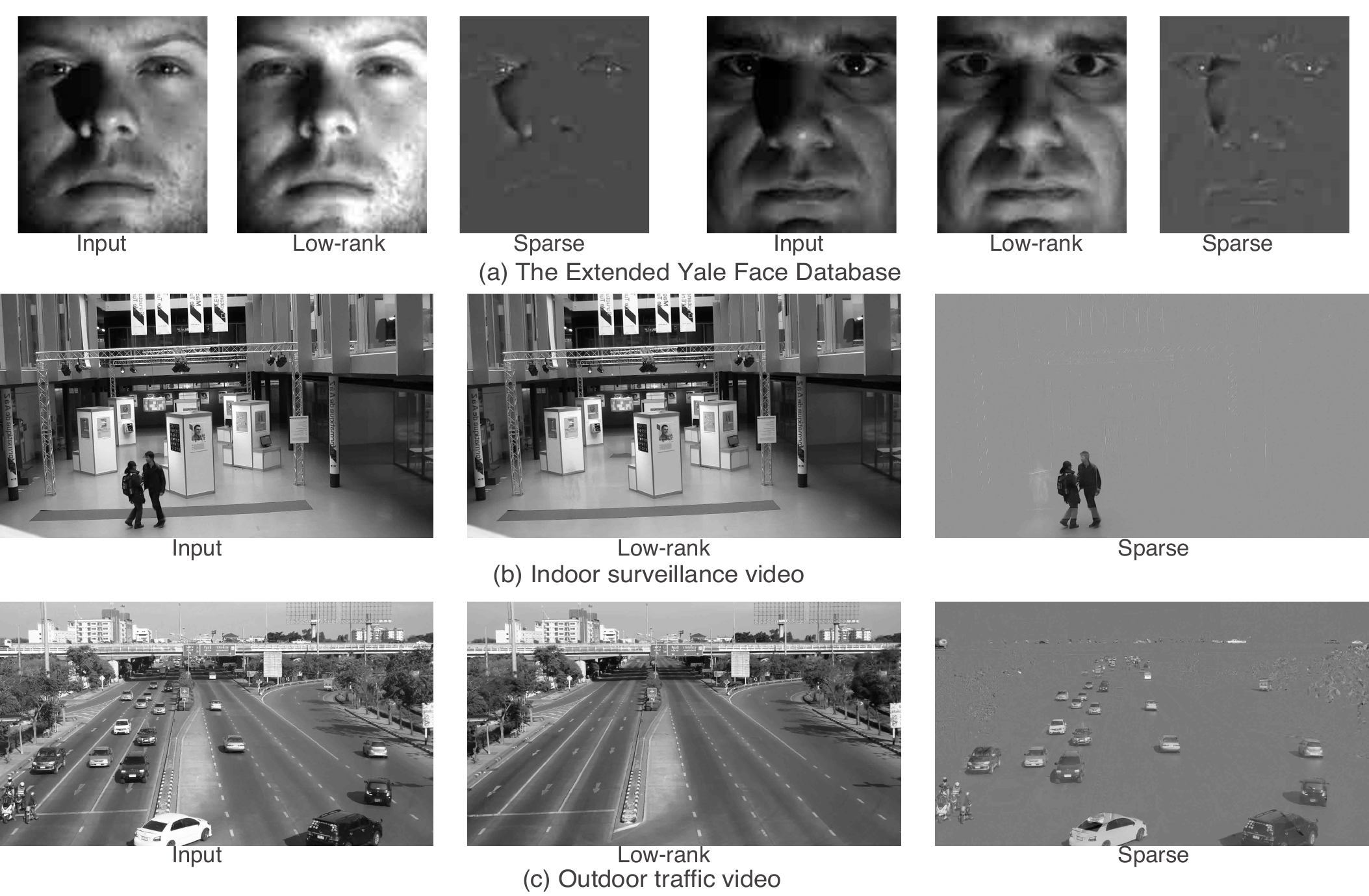}
 	\caption{Sample outputs of RPCA by RSVD on GPU.}
 	\label{fig:background}
 \end{figure}

\paragraph{Shadow removal.} 
We used the Extended Yale Database B~\cite{GeBeKr01} for shadow removal from face images shown in \fref{fig:background} (a). In the shadow removal experiment, $2383$ face images captured under various illumination conditions with a resolution of $168 \times 192 (=32256)$ were vectorized to form a $32256$-by-$2383$ input matrix. As shown in \Tref{tb:rpca}, our method using a GPU achieves moderate acceleration ($13.3 \times$) for this dataset.

\paragraph{Background subtraction.} 
We have used PEViD-UHD (Privacy Evaluation Ultra High Definition Video Dataset)~\cite{KorUHD2014} for assessing the background subtraction application shown in \fref{fig:background} (b). In this background subtraction experiment, $100$ video frames captured by a stationary indoor surveillance camera with a resolution of $1920 \times 1080 (=2073600)$ are used. They form a $2073600 \times 100$ tall-skinny input matrix. In this evaluation, it was observed that our method with a GPU exhibited $22.9\times$ speedup over a CPU implementation.
In addition, we used a Full-HD outdoor traffic video, which is heavily corrupted by camera jitterings and large traffic volume, indicating more corruptions (\fref{fig:background} (c). With a looser tolerance for evaluating convergence, \ie, the tolerance parameter $\mathrm{tol}$ was set to $10^{-5}$, our method achieved overall $24.3\times$ acceleration over a CPU implementation. 

Regarding the number of iterations in Algorithm~\ref{alg:rpca}, we found that there was a little difference in our method on GPU and a CPU implementation. This was caused by the slight accuracy difference between MKL and cuBLAS routines.

 
 \begin{table}[tb]
    
	\centering
	\caption{Comparison of RPCA by RSVD on CPU and GPU. All experiments are
		conducted in double precision. The parameters are set as: $k=p=10, q=1$. 
		The experimental setup is the same in \Sref{sec:experiments}.}
	\label{tb:rpca}
	\begin{adjustbox}{width=1\textwidth}
		\begin{tabular}{cccc|ccccc}
			
			Data &Image \#&Size (GB) & $tol$& & Iterations & RSVD time (s) & Total time (s)& Speedup\\
			\hline
			The Extended Yale Face &2383&0.57 &$10^{-7}$&CPU &30&4.0&26.6&-\\
			(168$\times$192) & & &$10^{-7}$ & GPU & 30&1.1 & 2.0 &13.3$\times$ \\  
			\hline
			Indoor surveillance & 100 & 1.5 & $10^{-7}$&CPU  & 30  & 50.6 & 121.4   & -\\
			(1920$\times$1080)& && $10^{-7}$&GPU & 30 & 3.4 & 5.3 &22.9$\times$\\
			\hline
			Outdoor traffic & 100 & 1.5 &$10^{-5}$ &CPU  & 25 & 51.2 & 104.7   & -\\
			(1920$\times$1080)& & &$10^{-5}$&GPU & 24 & 2.9 & 4.3  &24.3$\times$\\
			\hline
			
			\hline
		\end{tabular}
	\end{adjustbox}
\end{table}

\section{Conclusions}
\label{sec:conclusion}
Over the past several years, the SVD solution methods have shown significant advancement in terms of efficiency, but there have not been many attempts to harness the modern computing architectures such as GPUs. The likely explanation is that GPUs are still considered a specialized device; however, with that large-scale computations are now performed on cloud instances that are quite often equipped with GPU accelerators, developing new algorithms for new computing architectures is becoming urgent. This paper provides a fast randomized SVD algorithm that fully utilizes a GPU, that leads to state-of-the-art speed by a large margin. Our future work includes enabling BRSVD to run in a multi-GPU environment to achieve further acceleration.

\section*{Acknowledgments}
This research was supported by in part by the Japan Society for the Promotion of Science KAKENHI Grant Numbers 15K12008, 15H01687, 16H02801 and “Program for Leading Graduate Schools” of the Ministry of Education, Culture, Sports, Science and Technology, Japan.

\bibliography{RSVD}
\bibliographystyle{abbrv}

\end{document}